# Understanding and Modelling the Complexity of the Immune System:

# Systems Biology for Integration and Dynamical Reconstruction of Lymphocyte Multi-Scale Dynamics

Véronique Thomas-Vaslin

*CNRS, FRE3632, UPMC Univ Paris 06, UMRS 959, INSERM UMRS 959, Immunology-Immunopathology-Immunotherapy, Paris, France*

Correspondence: Dr. Véronique THOMAS-VASLIN UPMC- INSERM UMRS959, CNRS FRE3632, Immunology, Immunopathology, Immunotherapy, 83 Boulevard de l'Hôpital, Paris, 75013, France; Email: veronique.thomas-vaslin@upmc.fr

Understanding and modelling the complexity of the immune system is a challenge that is shared by the ImmunoComplexiT[1] thematic network from the RNSC

The immune system is a complex biological, adaptive, highly diversified, self-organized and degenerative cognitive network of entities, allowing for a robust and resilient system with emergent properties such as anamnestic responses and regulation. The adaptive immune system has evolved into a complex system of billions of highly diversified lymphocytes all interacting as a connective dynamic, multi-scale organised and distributed system, in order to collectively insure body and species preservation. The immune system is characterized by complexity at different levels: network organisation through fluid cell populations with inter- and intra-cell signalling, lymphocyte receptor diversity, cell clonotype selection and competition at cell level, migration and interaction inside the immunological tissues and fluid dissemination through the organism, homeostatic regulation while rapid adaptation to a changing environment.

Lymphocytes are the key actors of the immune system of vertebrates, in the middle of a multi-scale biological organization "from molecule to organism", and at the confluence with other different biological systems and the environment. Produced in central lymphoid organs by complex steps of differentiation, lymphocytes are then distributed in circulation and peripheral lymphoid organs to insure the integrity of the organism, composed of eukaryote and prokaryote cells and antigens. Each lymphocyte has a unique immuno-receptor type, generated by somatic gene rearrangement. The lymphocyte repertoire is able to recognize a potential repertoire of $10^{19}$ antigens. Lymphocyte repertoire selection is related to complex lymphocyte/Ag interaction to purge >95% of the cells that have not enough or to much affinity to antigens. It selects effector cells and regulatory cells involved in regulation of the immune response, insuring a dominant tolerance regulation process. There is a physiological basal production and activation of lymphocytes. The high turnover of lymphocytes and their immediate precursors is insured by daily renewal, differentiation and selection processes, proliferation and cell death, both in the central and peripheral lymphoid tissues. The cognition of antigens in individuals induces selection of lymphocyte repertoire via the immune-receptor. Thus, complex multilevel dynamic interactions at different time and space scales lead to a decentralised autonomous robust and resilient system. In young individuals this system is able to rapidly control pathologic situations, in most cases. However, with aging, perturbations occur at multi-scale levels, resulting in immune-depression (1).

---

[1] http://www.immunocomplexit.net/

From this general view of the immune system, a deeper understanding, quantification and modelling, from molecules to organism, of T cell differentiation, diversity, dynamics and repertoires selection are keys for fundamental research, medical advancement and drug discovery. A global evaluation is required to understand the "immune-physiome" multi-scale organisation, in order to evaluate inter and intra-individual variability/fluctuations but also the system's resilience that governs the emergence /immergence throughout biological scales. Studying the properties of the immune system with systems biology requires identifying entities, relations/interactions and processes through time. Experimental, mathematical and computer modelling interdisciplinary approaches are required to understand the system dynamics in physiological conditions and aging or during pathological perturbations or treatments. Despite recent systems biology initiatives to understand and model the immune system (2), we are still far from having the appropriate tools to understand its dynamics.

To explore lymphocyte dynamics from fundamental to clinical research, experimental methodologies and multiplex technologies should be adapted to capture the complexity. Depending on the species, various complementary approaches in vitro, ex vivo or in vivo might be used, to explore lymphocyte dynamics and turnover (3), (4) in health, aging and diseases (5). Then, analysis, integration and modelling of experimental multi-scale data also remain a challenge. The mining and analysis of these data requires both conventional reductionist and hypothesis driven approaches, completed by system biology that allows global investigation. These "bottom up" and "top-down" approaches complete each other and converge.

We will present our model based on multi-parameter and multi-scale experimental data obtained from normal healthy mice, describing the quantitative lymphocyte differentiation and selection, from thymocyte differentiation to peripheral effector/regulatory T cell immune response and feedback control. Our computer modelling software aims simulating and quantifying parameter values for production, cell fluxes, differentiation, proliferation, death, selection and homeostasis of CD4/CD8 T cells in physiological conditions, under the influence of different genetic origins. Our model also allows exploring the resilience of the system through aging or post perturbations as pathologies or treatments.

Traditional system dynamics models dealing with time are formalized with two distinct concepts as "continuous time" or "discrete time", by a succession of time points and intervals (6) even in "spatial" models. However, both type of models have limits (7). Models are often under-used because experimentalists can be reluctant to entertain mathematical formalism and because published models are largely disposable: rapidly forgotten after being published, instead of providing a foundation to build upon.
However, using visual language help to communicate and execute models. Immunologists often conceptualize the dynamical evolution of their systems in terms of "state-transitions" of biological objects and do it by means of personalized and informal graphical illustrations. Thus, the adoption of a more formal and standard type of state-transition diagram could improve the current situation not only to help biologists understanding better each other but also to ease the production and the reading of software code executing these visual transitions, at level of populations or agents (8). Organization of immune knowledge using a standardized, diagrammatic formal language should greatly improve knowledge integration at multi-scale levels and sharing between experimentalist and theoretician collaborators, rendering their software more readable, scalable and usable.
The Unified Modelling Language (UML) is a "High-level programming language" based on abstraction and using natural language that is easier to understand as compared to "low-level programming language", based on codes. Thus, visual modelling language considers biological-object as conceptual abstract-objects that endure processes. The level of abstraction allowed by these diagrams makes possible to distinguish more easily the "entities" as T-cells and the "processes" that occur at different levels such as cell differentiation, cell migration, cell interactions and cell cycle. Moreover, such "state-transition diagrams" allow computing parallel pathways at various scales to avoid redundancy that is inherent in the formal description of multi-level, heterogeneous and concurrent systems (9, 10). It also allows modelling heterogeneity in a very simplified and economical form (as compared to mathematical equations).

We present a "refactoring" of three published models of T-cell biology. Refactoring consists in restructuring the code or equations of a model to improve its expression, readability and extensibility, without changing its external behaviour. Two models consist in population-based model, with differential equations (continuous model) simulating cell differentiation first in the thymus (11) and then during activation /regulation during the course of an immune response (12). The other is an agent-based model simulating differentiation and migration of thymocytes moving as individual agents in space (13). Refactored models are now comparable (14), are directly executable and can provide simulations of physiology, pathologies and treatments.

Our experimental data and modelling reveal particular signatures of lymphocyte dynamics behaviour according to genetic origins and the robustness of the young immune system to transient perturbations. However, with aging production, differentiation, proliferation and T cell repertoire selection are affected, leading to increased perturbations and lower diversity index in the T cell repertoires, poor resilience of the system and increased inter-individual variability.